\documentclass[preprint1]{aastex631}

\usepackage{graphicx} 
\usepackage{natbib}
\usepackage{amsmath}
\usepackage{hyperref}
\usepackage{gensymb}
\usepackage{hyperref} 
\usepackage{enumitem} 
\usepackage{mathtools}
\usepackage{xcolor}
\usepackage{soul}

\hypersetup{
  citecolor=blue, 
}
\begin{document}

\title{A Robust Deep Learning Framework for Prominence Detection\\ through Composite Feature Representations}

\correspondingauthor{Harry Birch}
\email{harry.birch@northumbria.ac.uk}

\author[0009-0000-4625-1735]{Harry Birch}
\affiliation{School of Engineering, Physics, and Mathematics, Northumbria University, Newcastle upon Tyne, NE1 8ST, UK}

\author[0000-0001-8954-4183]{St\'ephane R\'egnier}
\affiliation{School of Engineering, Physics, and Mathematics, Northumbria University, Newcastle upon Tyne, NE1 8ST, UK}

\author[0000-0001-5678-9002]{Richard Morton}
\affiliation{School of Engineering, Physics, and Mathematics, Northumbria University, Newcastle upon Tyne, NE1 8ST, UK}

\begin{abstract}
Solar prominences are dynamic structures suspended within the solar corona and are manifestation of solar activity. Their evolution includes eruptions linked to coronal mass ejections, making their detection critical for space weather monitoring and forecasting. The vast amounts of high-cadence data provided by missions such as SDO/AIA motivate the application of deep learning frameworks capable of assimilating large-scale datasets. However, previous studies have reported poor model performance caused by contamination from hot coronal emission from the EUV He{\sc ii} 304~\AA\ channel. Using an existing labeled prominence dataset, we find that trained YOLOv5 object detection models exhibit a strong bias towards the 304~\AA\ colormap, rather than physically meaningful prominence features. We develop a further two models comprising three-channel images constructed through an original dataset preprocessing pipeline: (i) full-disk grayscale, full-disk enhanced corona, and disk-removed, (ii) same as (i) with all disk-removed images. Our pipeline corrects instrument degradation to maintain more consistent feature representations across the solar cycle. The composite model (i) achieves a mAP@50 of 0.749 and a recall of $78\%$ on the test set, outperforming previous bounding box methods. Visual analysis of the composite models reveals that many apparent false positives are valid unlabeled prominences. We additionally demonstrate cross-instrument generalization by testing the composite model on SUVI image data. By examining dataset biases that propagate into model predictions, we provide recommendations for robust dataset construction. We present a reliable, physically-motivated, and versatile deep learning model to automatically detect prominences in EUV images, providing a framework beneficial for space weather applications.
\end{abstract}

\keywords{Solar Prominences (1519); Convolutional neural networks (1938); Astronomy image processing (2306)}

\section{Introduction} \label{sec:Sect1} 
Solar prominences are structures of relatively cool plasma ($T \sim 10^{5}$~K) suspended in the hot solar corona ($T \sim 2\times10^{6}$~K), appearing as bright emission features in the He{\sc ii} 304~\AA\ pass-band located above the solar limb \citep{2010SSRv..151..243L,2010SSRv..151..333M, 2014LRSP...11....1P}. When viewed on the disk, they are termed `filaments', and instead appear as dark absorption features, tracing channels along the polarity inversion line (PIL). The behavior of prominences is intimately linked to other dynamic solar phenomena such as coronal mass ejections and flares; thus understanding their physics is important to the study of space weather and its effects at Earth. 

Historically, early detections of prominences were made through ground-based observations in visible wavelengths such as the Balmer H$\alpha$ spectral line, and then space-based with the emergence of satellite instrumentation. Modern instruments can capture many high-resolution images with a short cadence; the Atmospheric Imaging Assembly \citep[AIA;][]{2012SoPh..275...17L} onboard the Solar Dynamics Observatory \citep[SDO;][]{2012SoPh..275....3P} alone captures around 70,000 images per day. It is estimated that the amount of data collected by SDO/AIA is now in the petabytes. The vast amounts of data produced by these instruments has driven the need for processing methods that can handle large swaths of data. Many processing methods exist \citep[see][for a review]{2010SoPh..262..235A}, but machine-learning, and later deep-learning, methods are able to effectively harness such large amounts of data without the need of handcrafted rules. Early applications of machine learning to prominence detection include the study by \citet{2010SoPh..262..449L}, where the authors used a support vector machine (SVM) to classify prominences on the limb using Solar and Heliospheric Observatory / Extreme ultraviolet Imaging Telescope (SOHO/EIT) images in the 304~\AA\ channel. They were able to identify prominences and active region loop systems above the limb, although a prevailing problem was attempting to remove the diffuse emission from the background corona in the image. In the same year, \citet{2010ApJ...717..973W} showed an application of linear discriminant analysis (LDA), based on prominence shape and brightness, to create an automatic prominence detection method along with derived velocities applying to STEREO-B/SECCHI/EUVI data.

With the rapid evolution of deep-learning networks, numerous studies have sought to apply these techniques for the automatic detection of various solar features \citep[see the review paper by][]{2023LRSP...20....4A}. Much recent work has focused on detection of filaments on-disk \citep{2019SoPh..294..117Z,2022SoPh..297..104G,2024ApJ...965..150Z,2024A&A...686A.213D, 2025ApJ...980..176Z}, particularly through segmentation models such as U-Net \citep{ronneberger2015u}, which are designed for pixel-level delineation of features. There are fewer deep-learning studies of prominences, and to our knowledge,  no study focuses solely on prominence detection. \citet{2019SoPh..294...80A} constructed a convolutional neural network (CNN) trained on Hinode/SOT data to detect 5 solar classes: filaments, prominences, flare ribbons, sunspots and the quiet Sun. The model achieved $99.9\%$ accuracy on classifying unseen SOT images, however, struggled to identify any prominences from SDO/AIA in 304~\AA\ channel due to coronal emission. In the study by \citet{2021SoPh..296..160B}, the authors compared the performance of different detection models --- Single Shot MultiBox Detector \citep[SSD;][]{liu2016ssd} and Faster Region-based Convolutional Neural Network \citep[Faster R-CNN;][]{ren2015faster} --- for classification of coronal holes, sunspots, and prominences, using Helioseismic and Magnetic Imager (HMI) and AIA data. Once again, prominence detection showed relatively weak performance of around 50\% accuracy for both models. Recent advancements by \citet{2024ApJS..272....5Z} utilized a U-Net to construct an automatic detection system for prominences and active regions on the solar limb. Using tone-mappings to distinguish and contrast between the features, they detected over 50,000 prominences across a ten year period, obtaining statistics on solar cycle variations. Interestingly, the metric scores for active region detections were higher than prominences implying that prominences are harder to detect.

\medskip

A recurring challenge in detection studies of solar prominences is their inherent structural and dynamical variability, along with their low signal-to-background contrast. Compared to other solar structures, such as sunspots, coronal holes, and filaments, which can each be observed as isolated features with a strong contrast to their background, solar prominences represent a more challenging scenario for detection models when observed in the 304~\AA\ channel. The 304~\AA\ channel is most suitable in EUV for prominence observations due to the He \textsc{ii} emission at 304.8~\AA\ \citep{2010A&A...521A..21O} which corresponds to the cool prominence material. However, the additional sensitivity to the hotter Si \textsc{xi} 303.3~\AA\ emission line obscures much of the finer structural detail and is especially prominent in off-limb observations \citep{2000SoPh..195...45T, 2010A&A...521A..21O}. This presents a challenge for detection networks. Additionally, nearby active regions can further obstruct structural detail of prominences, presenting a problem both for labeling and detection. Therefore, prominences require greater care in constructing accurate detection systems. In the study by \citet{2024ApJS..272....5Z}, the authors employ different tone mappings to enhance the contrast of prominences and active regions against each other and the background. The response fitting method \citep[RFit;][]{2024SoPh..299...94A}, and more recently a deep-learning approach utilizing generative networks \citep[DeepFilter;][]{McMullan2026DeepFilter} are newly developed techniques for removing the hotter Si \textsc{xi} component from the 304~\AA\ channel. However, such methods increase the complexity of our pipeline. The wavelet-optimized whitening \citep[WOW;][]{2023A&A...670A..66A} technique presents a promising alternative, enhancing prominence structure, and bypassing the obscuring effects of the 304~\AA\ channel and active regions.

\medskip

The You Only Look Once \citep[YOLO;][]{redmon2016you,redmon2017yolo9000,2018arXiv180402767R} models are a series of deep neural-networks specializing in object detection applications, isolating features within a rectangular bounding box. Unlike two-stage detection models \citep[such as R-CNN][]{girshick2014rich}, which first generate region proposals before classifying them, YOLO's approach consists of a single detection stage, facilitating fast, efficient detection. The fifth iteration of YOLO, YOLOv5, was developed by Ultralytics \citep{yolov5} and offers strong PyTorch integration. Consequently, it has been widely adopted across computer vision applications. In solar physics, however, segmentation models remain the predominant choice for feature detection, and there are few studies employing YOLO. YOLOv5 was used in the study by \citet{2024A&A...686A.213D} as part of a larger pipeline to perform pixel-wise classification of solar filaments. However, to our knowledge, YOLOv5 has not been applied directly to prominence detection, where bounding box localization can be advantageous for features with ambiguous or diffuse boundaries, providing a different perspective from segmentation-based approaches.

\medskip

In this work, we apply and optimize the YOLOv5 network for prominence detection. We use the labeled dataset of full-disk SDO/AIA 304~\AA\ prominence observations created by \citet{2021SoPh..296..160B}. Through our first prominence model, we evaluate performance and uncover bias from the choice of colormap. This finding motivates a second approach using composite data comprising three channels (grayscale, WOW, and disk-removed) that enhance prominence structure and contrast. We compare the models through both statistical metrics and visualization, finding that the composite model performs best, achieving a recall of $78\%$ and mAP@50 of 0.749. The visual analysis additionally reveals that many apparent false positives are valid unlabeled prominences. We demonstrate that the composite model has strong generalization capability with cross-instrument applications by testing the model on data from the Solar Ultraviolet Imager \citep[SUVI;][] {2022SpWea..2003044D}. Finally, we discuss how dataset construction can introduce biases and provide guidelines to mitigate these effects.

The paper is organized as follows: Section~\ref{sec:Sect2} introduces the YOLOv5 model architecture, evaluation metrics, and the datasets used for the study. Section~\ref{sec:Sect3} covers the training and testing of the 304 colormap model along with the model biases. Section~\ref{sec:Sect4} details the results of the composite models. In Section~\ref{sec:Sect5}, we scrutinize model performance through test statistics and visual analysis, compare our results with previous prominence detection studies, demonstrate generalizability to SUVI data, and examine dataset biases. Section~\ref{sec:Sect6} presents a summary of our results, including recommendations for effective dataset construction, and future work involving our model.

\section{Method \& Dataset} \label{sec:Sect2} 
\subsection{Object Detection} \label{subsec:Subsect2.1}
Object detection models can be categorized by detection task: classification, box detection, and pixel-by-pixel segmentation. The latter two perform instance detection (detecting multiple objects within an image). Segmentation models, such as Mask R-CNN \citep{he2017mask} and the U-Net \citep{ronneberger2015u} family of models, have seen increasing popularity in solar physics applications \citep[see review by][]{2023LRSP...20....4A}. However, accurate segmentation labeling requires clearly defined feature boundaries, which presents a challenge for objects without distinct edges such as prominences. In a recent study, \citet{2021ApJ...913...28R} compared the results of several different automatic coronal hole detection schemes, finding a large discrepancy in location of the boundaries. Boundary ambiguity can be addressed through processing techniques \citep{2024ApJS..272....5Z}, semi-supervised hybrid approaches \citep{2024A&A...686A.213D}, or unsupervised methods \citep{2024ApJ...965..150Z}. However, these methods can often lead to increasingly complex pipelines. Alternatively, bounding box models like the Single Shot MultiBox Detector \citep[SSD;][]{liu2016ssd} require only the definition of a bounding box around each feature, without the need for complex or ambiguous boundary conditions. Such models can then localize objects within an image, assigning the most optimal bounding box around them. Nonetheless, these types of models are constrained by box orientation being restricted to the $xy$-plane, and inclusion of background regions that can impede effective feature learning. 

\medskip

The You Only Look Once \citep[YOLO;][]{redmon2016you,redmon2017yolo9000,2018arXiv180402767R} object detection system operates within the broader framework of convolutional neural networks (CNNs) for computer vision applications. More specifically, it belongs to the family of one-stage detectors (including SSD and RetinaNet) and treats object detection as a direct regression task. This contrasts with two-stage detectors, which first generate region proposals before classifying them. YOLO's approach leads to fast, efficient detections, by leveraging global image context for real-time object localization and classification.

While numerous variants and improvements to the YOLO architecture exist, we utilize YOLOv5\footnote{Github code available at \url{https://github.com/ultralytics/yolov5}.} \citep{yolov5} due to its longstanding support, strong PyTorch integration, extensive documentation, and widespread adoption in academic studies, which together support robustness and reproducibility. The incremental improvements offered by more recent models are beyond the scope of this work, as YOLOv5 provides performance sufficient for our application. The YOLOv5 architecture, like other object detection models, consists of three core components: a backbone for feature extraction; a neck for feature aggregation, and a YOLOv3 detection head which operates at three spatial scales for identifying objects of varying sizes \citep{yolov5}. YOLOv5 employs the SiLu  (Sigmoid Linear Unit, also known as Swish) activation function throughout the network and uses Stochastic Gradient Descent (SGD) with momentum as the default optimizer. Further details about the model hyperparameters are provided in Sections~\ref{sec:Sect3} and~\ref{sec:Sect4}.

\subsection{Performance Metrics}
Evaluation of machine learning model performance is typically achieved through the use of different metrics, the choice of which is dependent on the task at hand. For classification problems, values of true positives (TP, correct positive predictions), false positives (FP, false alarms), true negatives (TN, misses), and false negatives (FN, correct negative predictions), can be used to construct other meaningful metrics, in the form of the precision
\begin{equation}
\mathrm{Precision} = \frac{\mathrm{TP}}{\mathrm{TP} + \mathrm{FP}}~\textrm{,}
\end{equation}
the recall
\begin{equation}
\mathrm{Recall} = \frac{\mathrm{TP}}{\mathrm{TP} + \mathrm{FN}}~\textrm{,}
\end{equation}
and the F1 score
\begin{equation}
\mathrm{F1} = \frac{2\,\mathrm{TP}}{2\,\mathrm{TP} + \mathrm{FP} + \mathrm{FN}}~\textrm{,}
\end{equation}
which itself is the harmonic mean of the precision and recall. Integrating the precision as a function of recall provides the average precision (AP).

Another commonly used metric is the accuracy, given by the ratio of total correct predictions against the total of all predictions. However, in object detection, true negatives are ambiguous (there are no labels for regions without a feature), making accuracy uninformative. Instead, object detection performance is assessed through the precision, recall, along with summary metrics such as the mean average precision (mAP).

Before we can compute the mAP, it is necessary to establish a matching criterion to determine when a bounding box is considered a true positive. This is achieved through the Intersection over Union (IoU), which evaluates the correspondence between predicted and ground-truth areas \citep{rezatofighi2019generalized}. For areas A and B, the IoU is defined as
\begin{equation}
\mathrm{IoU}(A, B) = \frac{|A \cap B|}{|A \cup B|}.
\end{equation}
The mean average precision (mAP) is a standard benchmark metric widely used in object detection. It combines all of the above metrics, evaluating the area under the precision-recall curve for each class, where detections are considered correct if their IoU with the ground truth exceeds a certain threshold. The mean of this result is taken across all classes, combining classification and localization performance into a single, interpretable measure. A commonly used variant is the $mAP@50$, which is computed at an IoU threshold of 0.5 and written mathematically as
\begin{equation}
\mathrm{mAP@50} = \frac{1}{N} \sum_{k=1}^{N} \mathrm{AP}_{k}^{IoU=0.5}~\textrm{,}
\end{equation}
where $N$ is the number of classes. YOLOv5 widely reports the mAP@50 along with the mAP@50-95, which is taken over multiple IoU thresholds from 0.5--0.95 in 0.05 increments. These two metrics reflect detection difficulty: mAP@50 is more permissive, requiring only a $>50\%$ overlap for a successful prediction, whereas mAP@50-95 evaluates multiple IoU thresholds of increasing strictness up to $>95\%$ overlap. Thus, values of the mAP@50-95 are harder to optimize with results $\sim$0.5 being considered successful.

\subsection{Datasets} \label{subsec:Subsect2.2}
Many studies of prominences utilize observations in $H\alpha$, in which prominences appear as bright features strongly contrasted against the dark background at the limb. In EUV observations, the He \textsc{ii} resonance line at 304~\AA\ which has a peak emission temperature around 80,000 K, similarly provides strongly contrasted prominence observations, along with insight into the prominence-corona-transition region (PCTR) due to the higher temperature sensitivity \citep[see review by][]{2014LRSP...11....1P}. We use full disk observations taken using the 304~\AA\ channel from the Atmospheric Imaging Assembly \citep[AIA;][]{2012SoPh..275...17L} onboard the Solar Dynamics Observatory \citep[SDO;][]{2012SoPh..275....3P}. These EUV observations offer several advantages relative to ground-based $H\alpha$ observations, including resistance to atmospheric absorption due to their space-based nature, near-continuous full-disk coverage, and simultaneous multi-thermal diagnostics across multiple channels sensitive to different coronal temperatures. Other SDO/AIA EUV channels, such as the 171~\AA\ and 193~\AA\ channels, can provide additional information about the prominence environment but cannot capture the full detail of the cool prominence material visible in 304~\AA. However, since the launch of SDO in 2010, the CCD camera of the 304~\AA\ channel has been subject to a noticeable degradation, which has reduced the dynamic range of intensity detection.

In this study we implement the dataset of labeled prominences images\footnote{Data accessed from \url{http://sdo.kasi.re.kr/dataset_object_detection.aspx.}} created by \citet{2021SoPh..296..160B}. These are a series of full disk observations taken using the 304~\AA\ channel from SDO/AIA, of size $512\times512$ and in the JPEG format. The temporal range of the data is from 2010--2017 with a cadence of 4 hours. However, data from later years are sparser, likely because prominences are less clearly visible in these degraded images and therefore were not labeled. The label files where converted in the YOLOv5 format from XML type using the ROBOFLOW software \citep{Dwyer2024}. The label files contain the class ID of each box, along with the normalized coordinates for the box’s center ($x$,$y$), width, and height; therefore, they are invariant to image size. We use a split of 70:20:10 for training, validation, and testing sets, respectively. There is a total of 2924 images in our dataset\footnote{Two images were removed from the original dataset due residual roll misalignment - these images were not correctly rotated to have the $y$-axis align with the solar North. A further 6 images were found to erroneously have empty label files, and were subsequently re-labeled.}, with a total of 9236 prominence instances. We will henceforth refer to this labeled dataset as the ground truth in our analyses, and the model trained on this data as the 304 colormap model.

\medskip

For Section~\ref{sec:Sect4}, we use a improved version of this dataset, motivated by limitations exhibited by the first model which arose due to biases in the \citeauthor{2021SoPh..296..160B} dataset. As discussed previously, prominences observed in the 304~\AA\ channel can often be challenging to identify due to the hotter Si \textsc{xi} emission. Additionally, bright features such as active regions may obscure prominences. In the approach by \citet{2024ApJS..272....5Z}, different tone-mappings were adopted to discriminate between prominences, active regions, and the background. To visually enhance prominences, we use the wavelet-optimized whitening \citep[WOW;][]{2023A&A...670A..66A} technique. This approach allows for the enhancement of fine-structural detail whilst also preserving the large-scale structure. In additional, its edge-aware variant incorporates bilateral filtering, which preserves edges and boundaries. The denoising coefficients operate across multiple scales, applying adaptive thresholds that suppress noise while retaining genuine small-scale structures. Compared to other processing methods, WOW provides stronger contrast and greater computational efficiency, making it well suited for application to large datasets such as ours. 

\begin{figure*}[htbp!]
\graphicspath{ {./images/} }
    \centering
    \includegraphics[width=0.75\linewidth]{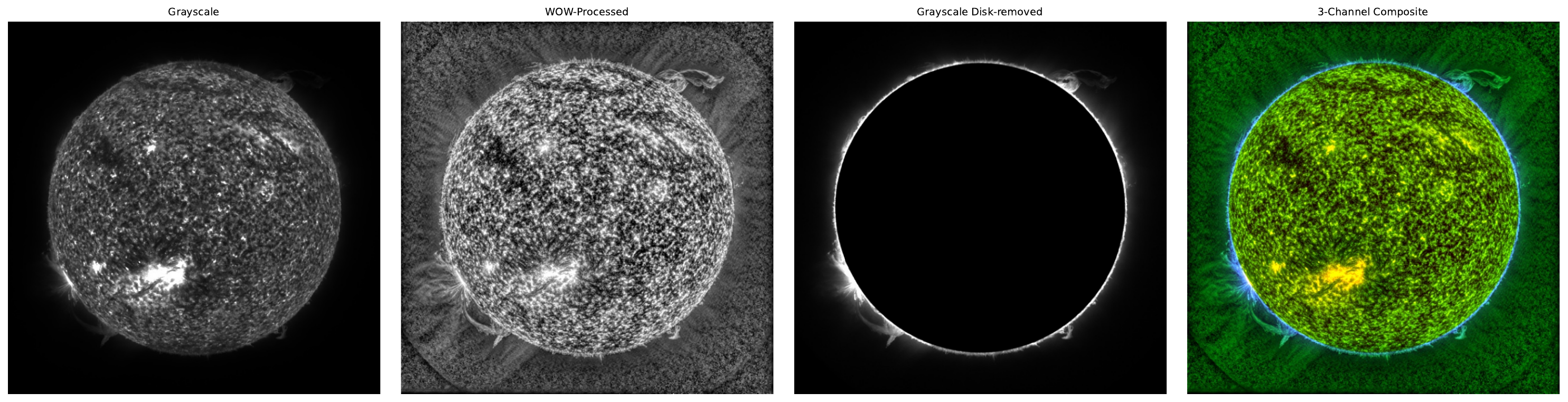} 
    \caption{Examples of the data in each channel, shown from left to right: grayscale, WOW-processed, disk-removed grayscale, and the 3-channel composite image formed by combining all three channels.}
    \label{fig:Fig1}
\end{figure*}

We first calibrated the FITS data to level 1.5, applying corrections for instrument degradation, before resizing down to an image size of $640\times640$ (the default input size for YOLOv5). We apply the edge-aware denoised WOW transform on the resized image array, with parameters for the bilateral filter of 1 and denoise coefficients [5,2]. These values provide a balance between noise suppression and preservation of fine-scale structure. The WOW transform takes less than a second per resized image, and achieves comparable results to transforms applied to 4K images followed by downscaling. The resulting array is percentile clipped (0.5--99.5) using values from the first image in the sequence, and rescaled to the range 0--255, an example of which is shown in Figure~\ref{fig:Fig1}. This normalization procedure ensures that later, more degraded images, retain the same contrast as the first images in the dataset.

A by-product of the WOW transform is amplification of noise off the limb. Whilst denoising can suppress this effect, some residual noise will always be present. Therefore, we include further channels to our input tensor with different representations of the prominence data. The inclusion of more channels can help model invariance if these other channels capture different representations of the data (increased feature diversity). This can also have the adverse effect of encouraging overfitting and increasing the computational cost. We include two further channels: the first being a grayscale representation of the original FITS data array, and the second grayscale representation with the disk removed, shown by the first and third panels of Figure~\ref{fig:Fig1}, respectively. These channels are processed following the same calibration and preprocessing pipeline as the WOW channel, including degradation correction, downscaling, and normalization to 0–255. The grayscale channel contains an original, unmodified representation of the FITS data array, whilst the disk-removed channel blocks out the solar disk to enhance prominences. 

The composite dataset combines these three results to generate a tensor of dimension [640, 640, 3], which will serve as the input to our model in Section~\ref{sec:Sect4}, as seen in the rightmost panel of Figure~\ref{fig:Fig1}, with the order of channels being grayscale, WOW-processed, and disk-removed. For comparison, we also created a composite disk-removed dataset in which all three channels contain data representations with the disk removed. In this case, the first channel was additionally log-scaled to distinguish it from the third channel with the disk-removed. We use the same data split of 70:20:10 for training, validation, and testing sets, respectively. To enable direct comparison between the models, we are using the same random seed as for the 304 colormap model dataset. Models trained on these datasets are referred to as the composite models, with the composite disk-removed model denoting the variant without the solar disk.

\section{304 Colormap Model} \label{sec:Sect3} 
\subsection{Model Training and Hyperparameters} \label{subsec:Sect3.1}

The main training hyperparameters for the 304 colormap model are reported in the first column of Table~\ref{tab:Tab1}. We train the model from scratch for 300 epochs with early stopping; tests employing pretrained weights yielded no improvement over training from scratch. On-the-fly data augmentations were used during training to increase the diversity of the training dataset. The corresponding augmentation parameters are listed in the first column of Table~\ref{tab:Tab2}. We additionally included modifications to the augmentation code for increased suitability, changing the rotation padding color in the \texttt{degrees} parameter to black to match the colormap background, and limiting the scaling range to $[1-\texttt{scale},1]$ (asymmetric scaling). These augmentations, along with the \texttt{flipud} (up/down) and \texttt{fliplr} (left/right) parameters, help to increase model robustness and generalization to other instrument datasets (see Section~\ref{subsec:Subsect5.3} on SUVI data). Lastly, we chose a \texttt{hsv\_h} modification of 0.5 for the hue to reduce the prediction dependency of the 304 colormap. All other data augmentation hyperparameters are set to zero, and the remaining training hyperparameters are left at their default values.

\begin{table*}[htbp!]
\centering
\caption{Model training hyperparameters for all models\footnote{All models were trained using a A100 GPU.}}
\label{tab:Tab1}
\begin{tabular}{c c c c}
\hline
\hline
Hyperparameter & 304 Colormap Model & Composite Model & Composite Disk-removed Model \\
\hline
Model Scale        & yolov5m (medium) & yolov5s (small) & yolov5s (small) \\
Epochs (early stopping) & 300 (289) & 300 (220) & 300 (293) \\
Batch Size   & 32 & 32 & 32 \\
\hline
\end{tabular}
\end{table*}

\begin{table*}[htbp!]
\centering
\caption{Augmentation hyperparameters for all models}
\label{tab:Tab2}
\begin{tabular}{c c c c}
\hline
\hline
Hyperparameter & 304 Colormap Model & Composite Model & Composite Disk-removed Model \\
\hline
hsv\_h (hue)            & 0.5 & 0 & 0 \\
degrees (degrees)      & $\pm 15$ & $\pm 15$ & $\pm 15$ \\
scale (range)          & [0.7, 1.0] & [0.7, 1.0] & [0.7, 1.0] \\
flipud (probability)   & 0.5 & 0.5 & 0.5 \\
fliplr (probability)   & 0.5 & 0.5 & 0.5 \\
\hline
\end{tabular}
\end{table*}

The model training results showed a mAP@50 score of 0.755 and a mAP@50-95 score of 0.36, with the best results observed at epoch 148. The output confusion matrix reported that $76\%$ of prominences where successfully detected in the validation set. Inspection of the F1-confidence curve shows a $\textnormal{F}1_{peak}$ score of 0.74 at a confidence threshold of 0.373, suggesting that the model performs best when the confidence threshold for predictions is low.

For additional analysis, we run model inference on the whole testing set, using a confidence of 0.40\footnote{Although the best confidence value for this model is given as 0.285, we choose the default value of 0.4 for consistent comparison with later models} and non-maximum suppression (NMS) IoU threshold of 0.45. In other words, boxes of confidence $>0.4$ are filtered-out, and the remaining boxes passed through the NMS function, which suppresses overlapping boxes based on the given IoU threshold. This is a strategy to reduce duplicate boxes (false positives). We apply an additional IoU threshold of 0.5 on the detections to generate the statistics, discarding predictions where the IoU between the ground truth and predicted boxes falls below this value.

\begin{table*}[htbp!]
\centering
\caption{Detection statistics for the different model configurations}
\label{tab:Tab3}
\begin{tabular}{c c c c c c c c c c c}
\hline
\hline
Model & TP & FN & FP & \# GT & \# Pred. & Precision & Recall & F1 Score & mAP@50 & mAP@50-95 \\
\hline
304 colormap & 712 & 239 & 339 & 951 & 1051 & 0.678 & 0.749 & 0.711 & 0.755 & 0.360 \\
Composite & 741 & 210 & 372 & 951 & 1113 & 0.666 & 0.779 & 0.718 & 0.749 & 0.351 \\
Composite Disk-removed & 707 & 244 & 340 & 951 & 1047 & 0.675 & 0.743 & 0.708 & 0.737 & 0.348 \\
\hline
\end{tabular}
\end{table*}

The first row of Table~\ref{tab:Tab3} lists the test statistics from inference, showing the number of true positives (TP), false negatives (FN), false positives (FP), ground truth boxes (\# GT), prediction boxes (\# Pred.), and performance metrics. The model correctly identifies $75\%$ of ground-truth events, indicating good recall. Similarly, $68\%$ of the predicted events correspond to true positives, reflecting moderate precision. Together, these yield an F1 score of 0.71, demonstrating a well-balanced detection performance.

\subsection{Model Bias} \label{subsec:Subsect3.2}
The 304 colormap model has been shown to perform well, despite the limited processing employed on the initial data. However, the choice of colormap for the images leads to concern over potential dataset bias. While a range of interpretability techniques exist, it remains non-trivial to determine which features a CNN learns and how these contribute to its final predictions. In their \citeyear{zeiler2014visualizing} seminal paper, \citeauthor{zeiler2014visualizing} visually unmask the filters at different layers of the network, showing that with increasing depth the filters learn increasingly complex features. Subsequent studies have investigated how different feature properties, such as shape, texture, and color, can influence the network predictions \citep{10.1371/journal.pcbi.1006613,geirhos2018imagenet,hosseini2018assessing,2020arXiv201206917S}. In particular, the feature cues exploited within the network filters are often determined by the target class --- for example animals show strong texture dependencies \citep{10.1371/journal.pcbi.1006613,geirhos2018imagenet}, whereas text-based detections rely more on shape \citep{hosseini2018assessing}. Likewise, color can also influence filter predictions, taking priority over shape and texture cues \citep{2020arXiv201206917S}.

\begin{figure*}[h]
\graphicspath{ {./images/} }
    \centering
    \includegraphics[width=\linewidth, trim=30 10 30 30, clip]{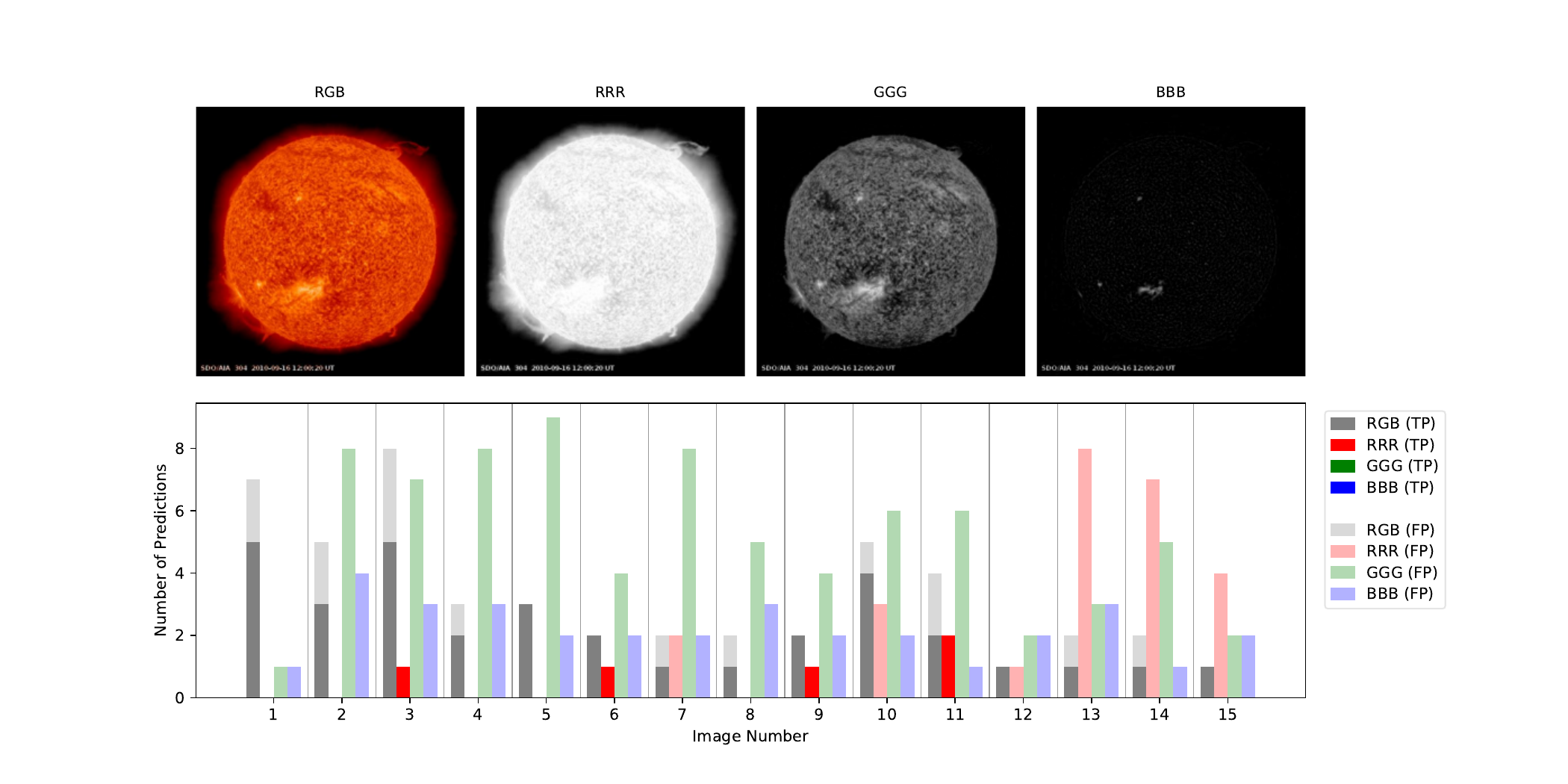} 
    \caption{Top: Example images for testing color bias showing (from left to right) the original RGB channel, and the three replicated channels (RRR, GGG, BBB). Bottom: Bar chart showing the number of predictions per image when testing inference on each of the four image types above. The bars are segmented between true positives (darker shades) and false positives (lighter shades).}
    \label{fig:Fig2}
\end{figure*}

To test for potential color bias we follow a similar method as employed by \citet{2020arXiv201206917S}, where incongruent images are created by modifications of the image channels: images are formatted using the RGB color model (Red, Green, Blue). We first test by isolating each channel, and replicating across all three channels (e.g. GGG), thereby constructing an image containing information from only one channel. This tests for a specific channel dependency. Examples of these replicate-channel images are presented in the top row of Figure~\ref{fig:Fig2} along with the original image (leftmost panel). The bottom plot shows the number of predictions per image, for inference on each of the image types in the top row. The bars are stacked according to true positives (darker shades) and false positives (lighter shades). The 15 images were randomly selected from the test set and ordered by time.

Deconstructing the RGB channels in this way highlights how the image data have been split across the different color channels. Predictably, the red channels contains most of the information both on disk and above the limb, and the prominences can be clearly observed. The green channel is comprised of mostly the disk and higher intensity areas which appear yellow in the original RGB image. We see less low intensity emission off the limb, as well as prominence material. Only the brightest regions are preserved in the blue channel, along with faint outlines of the remaining structures in the image.

In the bar chart below, the RRR images are the only replicated variant to score any true positive predictions. Interestingly, there is a marked rise in the number of false positive predictions in more recent images where the degradation has strongly effected image quality. This suggests that the 304 colormap model struggles detecting on degraded images. No true positives could be detected on the GGG and BBB type images, with the former routinely scoring the largest number of false positives predictions of all four image types. This is consistent with the visual information represented in these channels, which is insufficient for prominence detection.

\begin{figure*}[h]
\graphicspath{ {./images/} }
    \centering
    \includegraphics[width=\linewidth, trim=30 10 30 30, clip]{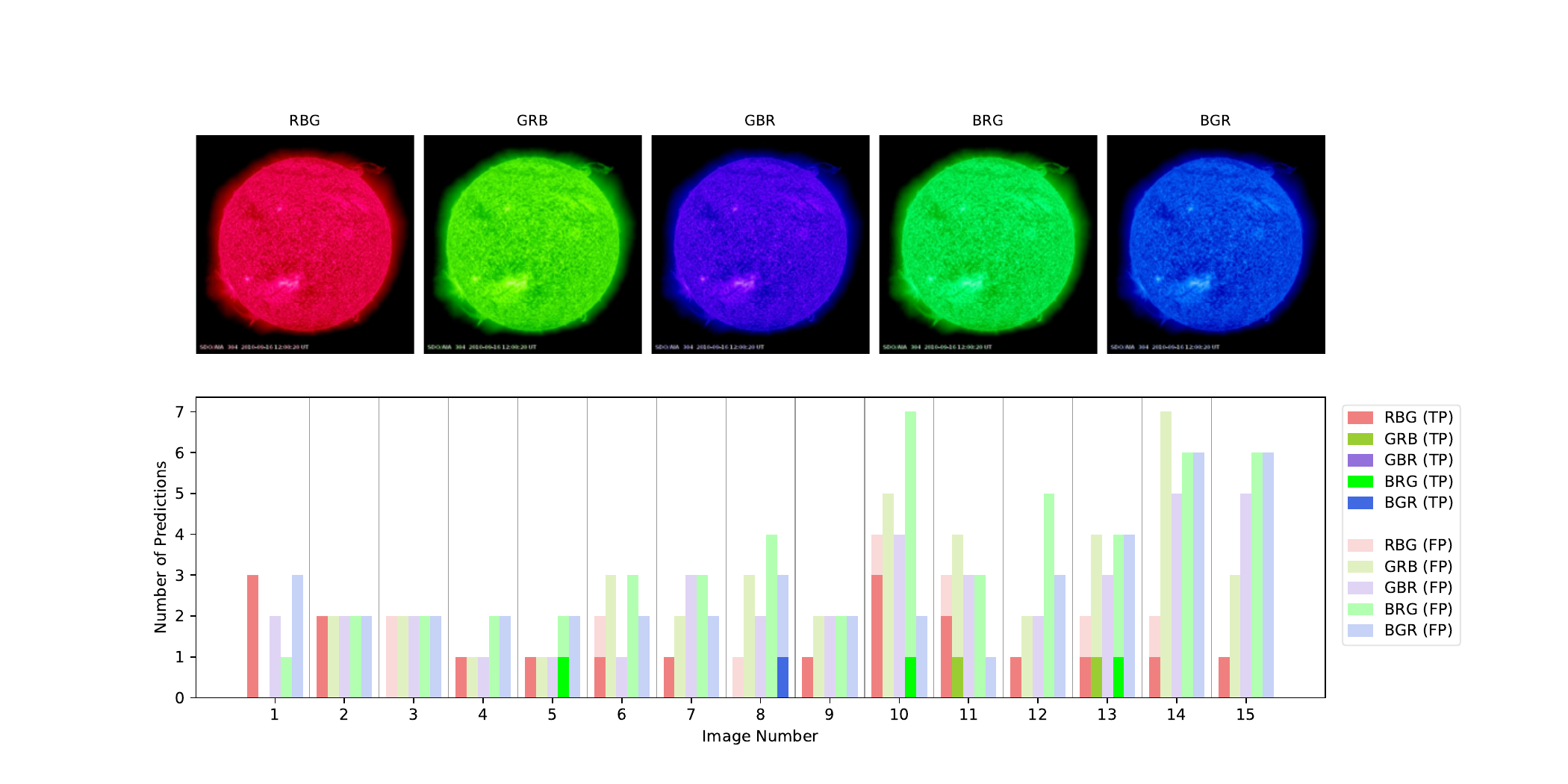} 
    \caption{Top: Example images for testing color bias showing the different RGB channel permutation (from left to right), RBG, GRB, GBR, BRG, BGR. Bottom: Bar chart showing the number of predictions per image when testing inference on each of the five permutation types above. The bars are segmented between true positives (darker shades) and false positives (lighter shades).}
    \label{fig:Fig3}
\end{figure*}

For a second test, we utilized permutations of the RGB channels to create incongruent images, thereby testing dependency on channel order. In addition to the default RGB channel order, there are five permutations in total (RBG, GRB, GBR, BRG, BGR), as shown in the top panel of Figure~\ref{fig:Fig3}. The bottom panel shows the number of predictions per image for inference on each of the image types in the top row. The bars are stacked according to true positives (darker shades) and false positives (lighter shades). We use the same sample of 15 test-set images as for the first test, and order the images by time.

The strongest true positives are made when the red channel is unmoved. True positives can also be seen in cases where the first two channels have the most information e.g. GRB, BRG, and BGR. The GBR is the weakest permutation, as no channels are in their original positions and there is only weak information in the first two channels. These results show a prioritization of channel position - the model expects most information to be in the first (red) channel, and then the second (green) channel. As with the first test, the number of false positive predictions increases for later images, suggesting influence of image degradation.

\section{Composite Models} \label{sec:Sect4} 
In order to control the bias present within the 304 colormap model, we move away from the original JPG format dataset of \citet{2021SoPh..296..160B} and the associated fundamental color tables for SDO/AIA images. We therefore employ the composite dataset, which was constructed from the original FITs level data, allowing for calibration, corrections for instrumental degradation, and additional image transforms to improve prominence contrast off the solar limb. We consider two variants of this dataset. The first comprises three channels corresponding to grayscale, WOW-processed, and disk-removed representations. The second is a disk-removed composite in which all three channels contain disk-removed data representations.

 The main hyperparameters for the composite models are shown in second and third columns of Table~\ref{tab:Tab1}. Unlike the 304 colormap model, we opt to use the smaller yolov5s model architecture, as tests revealed that this model generated more valid predictions (see Section \ref{subsec:Subsect5.1}) compared to the more conservative medium model. The rigid data preparation for the composite models renders any pre-training unnecessary, as the pre-trained features are now incompatible with our revised data distribution. Specifically, our new dataset contains no meaningful color information (colors are synthetic and carry no physical meaning) whereas the pre-training images encode inherent color dependencies. Likewise, color augmentations such as hue alterations will destroy the careful pre-processing we have constructed and therefore are unsuitable here. The remaining training augmentation parameters were unmodified from their previous values and are reported in the second and third columns of Table~\ref{tab:Tab2}.

\medskip

Training results for the composite model reported a mAP@50 score of 0.749 and a mAP@50-95 score of 0.351, with the best results from epoch 107. $80\%$ of prominences in the validation set were correctly identified. The F1-confidence curve peaked at $\textnormal{F}1_{peak}=0.73$ and a confidence of 0.507, indicating optimal performance is achieved at an intermediate confidence level.

The composite disk-removed model had slightly lower performance with a mAP@50 score of 0.737 and a mAP@50-95 score of 0.348. $74\%$ of the prominences in the validation set were successfully detected and the best epoch was 156. The F1-confidence curve peaked at a $\textnormal{F}1_{peak}$ score of 0.72 and confidence 0.421, showing strong performance with intermediate--low confidence. 

Following the testing pipeline used for the 304 colormap model, we repeat the analysis on the two composite models, running inference across the 292 images in the test set. These results are reported in rows two and three of Table~\ref{tab:Tab3}. Interestingly, the composite model outperforms the composite disk-removed model, which is the weakest of the three models tested. The composite model produces the largest number of detections, but has the greatest number of false positives and subsequently the lowest precision of $67\%$. However, the higher recall of $78\%$ suggests that this model scored best at correctly identifying the most ground-truth cases. The composite model with the disk-removed is more conservative in its predictions and has the lowest number of true positives of the three models; consequently, it exhibits the poorest overall performance.

\section{Discussion} \label{sec:Sect5} 
\subsection{Model Comparisons} \label{subsec:Subsect5.1}
We have constructed three models for prominence detection: the 304 colormap model, and two composite models. The motivation for the composite models follows on from the discovery of color bias present within the 304 colormap model. Tests uncovered variable performance suggesting that the model is overly sensitive to the red channel, implying a channel and positional dependency. As a result, we have demonstrated a bias towards features present in the red channel, indicating that the learned shape and textural cues are primarily constrained to this channel. Metric scores from Table~\ref{tab:Tab3} show similar performance between the three models, making it unclear from the metrics which model exhibits best detection performance. The composite model has the overall highest F1 score and recall, although it produces the most number of false positive predictions. The 304 colormap model achieves better training metrics and has the highest precision. The composite disk-removed model has the lower metric scores of the three. To further discriminate between the models we will focus on a visual comparison of these results.

\medskip

\begin{figure*}[ht]
\graphicspath{ {./images/} }
    \centering
    \includegraphics[width=0.75\linewidth]{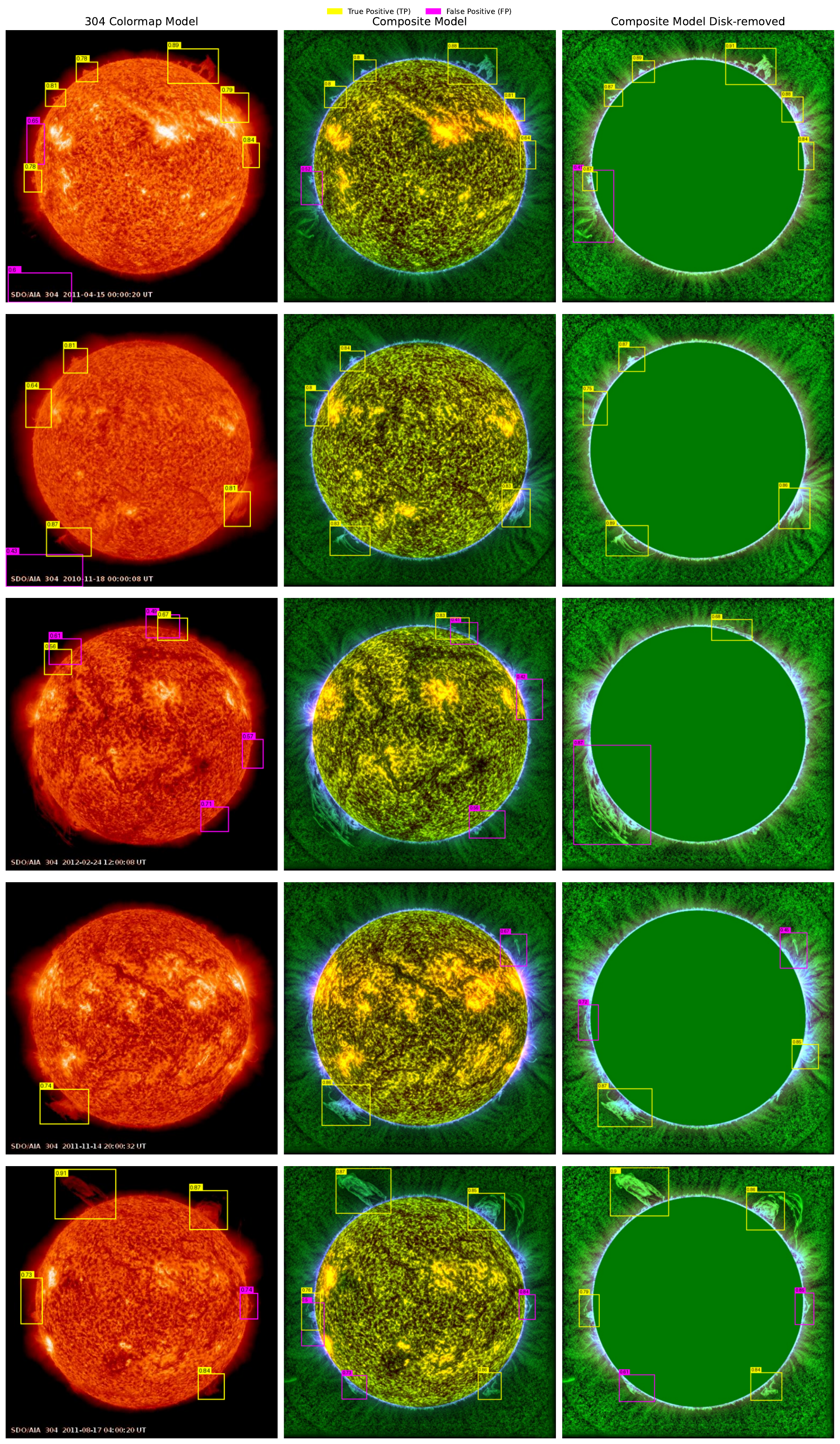} 
    \caption{Sample of inference results for the 304-colormap model (right), 3-channel composite model, and the composite models disk-removed. Boxes are colored according to true positives, in yellow, and false positives, in magenta. Confidence scores are displayed in the top-left corner of each box.}
    \label{fig:Fig4}
\end{figure*}

To verify the predictive performance of each model, we plot a random sample of the testing results in Figure~\ref{fig:Fig4}, with true positive predictions shown in the yellow boxes and false positives in the magenta boxes. Confidence scores are shown in the top-left corner of each box. The 304 colormap model effectively identifies many prominences despite the contribution of hotter coronal emission present in the 304~\AA\ channel observations. However, it is prone to spurious detections and, in two cases, incorrectly identifies overlaid date-time text (artifacts of the training data from \citeauthor{2021SoPh..296..160B}) as a prominence. 

The composite models both exhibit strong detection performance, more so than initially implied by the metrics. WOW processing in these models reveals the true extent of the prominences. For example, the high-latitude prominences in the bottom row of Figure~\ref{fig:Fig4} are only fully visible through WOW processing. However, the influence of the box size in the training labels have primed the model to predict only the portion of the prominence observable with the 304 colormap rather than the full extent provided by the WOW processing. Both models make more false positive predictions which, upon inspection, are often valid prominences. Distinguishing between the two composite models is challenging as both models capture prominences not seen by the other. In this sample, the disk-removed model displays higher precision, with all but one false positive being a valid prominence. The composite model is more exploratory, making more predictions but with lower precision.

In light of the visual analysis, the composite models appear strongest, with the disk-removed model the most consistent. This is despite the disk-removed model having the lowest metric scores. The detection metrics are misleading due to the labeling of the data, which was performed using the same 304~\AA\ channel images used for the 304 colormap model. As a result, the incomplete labeling of the ground-truth dataset introduces bias in the statistical metrics, as the labeling does not include prominences made visible through the WOW processing. These additional features are subsequently assigned a lower IoU against the ground truth. Furthermore, the labeling is incomplete; therefore, recall is more reliable than precision, as many apparent false positives are actually unlabeled true prominences. This is a clear example of how bias in the labeling process propagates directly into performance statistics. Based on a visual inspection of a sample, we estimate that of the false positive predictions in the composite model, approximately $74\%$ are actually valid prominences. The remaining $26\%$ consists of duplicate predictions, misclassifications of flare loop structures, and features where no prominence can be clearly distinguished. For the composite no-disk model, $80\%$ of false positives were valid prominences, with the majority of actual false positives resulting from duplicate boxes. Accounting for this, we determine the effective precision to be $91\%$ for the composite model and $93\%$ for the composite disk-removed model, as shown in Table~\ref{tab:Tab4}.

\begin{table}[h]
\centering
\caption{Effective detection statistics for the composite models.}
\label{tab:Tab4}
\begin{tabular}{lccc}
\hline
\hline
Model & Eff. TP & Eff. FP & Eff. Precision \\
\hline
Composite & 1016 & 97 & 0.913 \\
Disk-removed & 979 & 68 & 0.935 \\
\hline
\end{tabular}
\end{table}

Although the disk-removed composite model arguably achieves the strongest visual performance, we select  the composite model for further analysis due to its exploratory nature and highest recall. Counterintuitively, removing the disk in the composite model does not significantly improve performance despite prominences being easier to visually identify with this processing. Enhanced noise from log-scaling may have negatively impacted the model's predictive capability, increasing confusion. The training results further highlight problems with the disk-removed model, as slower convergence, evidenced by the later stopping time, indicates that model struggles to learn robust and generalizable features. By removing the disk, we have created a more challenging problem where the model has fewer contextual cues to make prominence predictions, despite having a simpler input representation.

\subsection{Comparison with Literature} \label{subsec:Subsect5.2}
Our composite model (recall: $78\%$, mAP@50: 0.749) substantially outperforms previous bounding box studies, improving 19-26 percentage points over the SSD ($59\%$) and Faster R-CNN ($52\%$) results from \citet{2021SoPh..296..160B} on the same dataset. This improvement reflects the advantages of both the YOLOv5 network and our multi-channel composite approach, where each channel contains physically relevant prominence representations. These results are also an improvement over the multi-class CNN model by \citet{2019SoPh..294...80A}, which struggled with prominence classification with SDO/AIA 304~\AA\ images, missclassifying prominences as flare ribbons due to coronal background emission. By focusing exclusively on prominence detection our model avoids this source of confusion. Furthermore, the WOW processing enhances prominence structures while suppressing diffuse coronal emission, directly addressing this issue. Detailed comparison with segmentation studies is complicated due to the different task formulation. \citet{2024ApJS..272....5Z} reported a recall of $88.4\%$ and a precision of $66.8\%$ using U-Net with tone-mapping pre-processing, compared to our $78\%$ and $67\%$, respectively. However, visual analysis suggests approximately $74\%$ of our false positives are unlabeled prominences, implying an effective precision is $91\%$ under complete labeling.

\subsection{SUVI Detection} \label{subsec:Subsect5.3}
\begin{figure*}[ht]
\graphicspath{ {./images/} }
    \centering
    \includegraphics[width=0.75\linewidth]{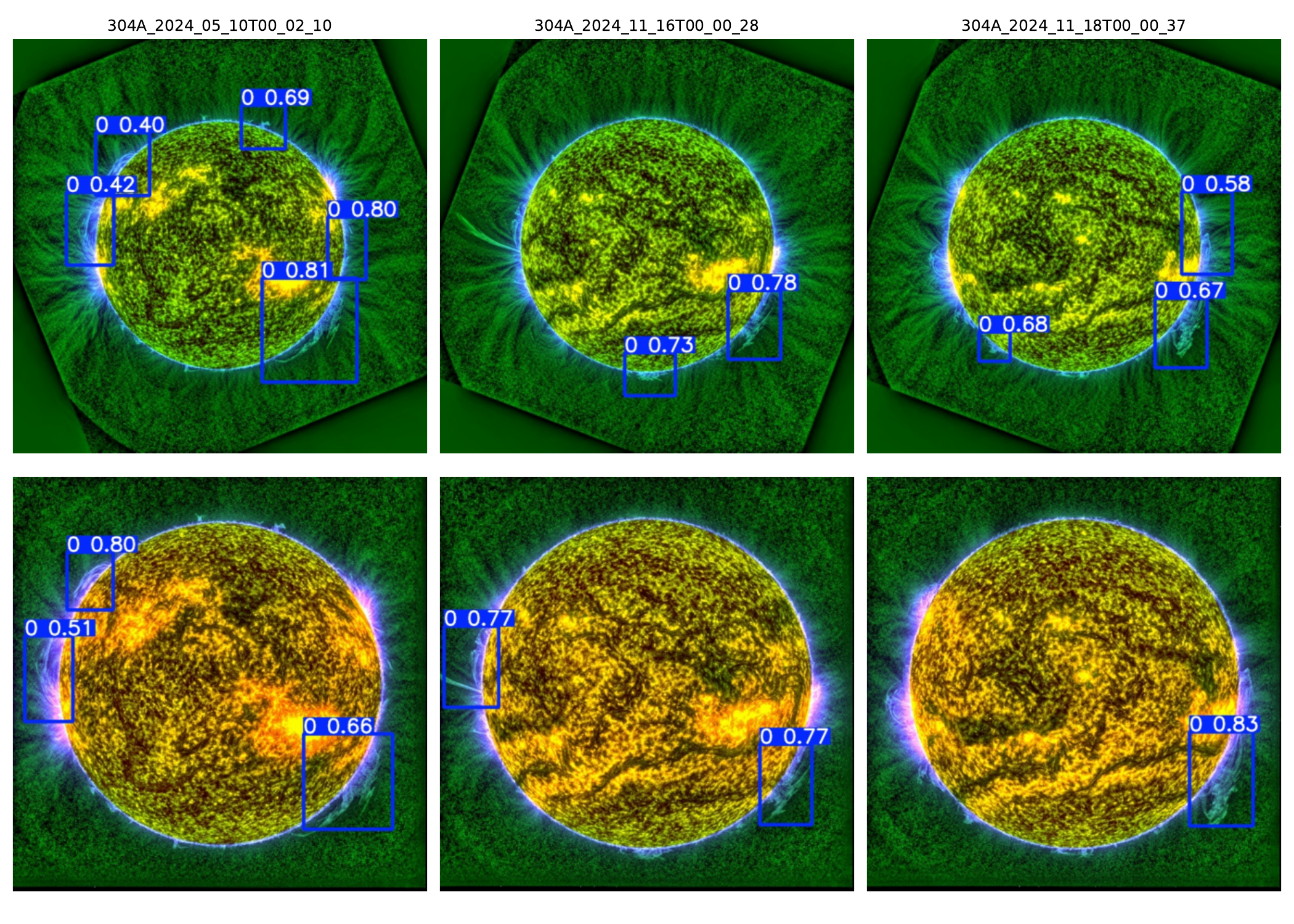} 
    \caption{Sample of inference results using the composite model testing on SUVI data (first row), and temporally matched SDO/AIA images (bottom row). Detection boxes are in blue with confidence scores displayed in the top-left corner of each box.}
    \label{fig:Fig5}
\end{figure*}

To evaluate model generalization, we perform inference on a set of images from the Solar Ultraviolet Imager \citep[SUVI;][] {2022SpWea..2003044D}, onboard the GOES-R series satellites. Data were collected between 1 March 2024 and 1 March 2025 for a total of 272 images, using the GOES-18 satellite. A sample of these inference results using the composite model is shown in top row of Figure~\ref{fig:Fig5}. For comparison, we show inference results on temporally matched AIA images in the bottom row.

The composite model successfully generalizes to SUVI images, despite the image rotation and larger field of view. The AIA images show agreement; five of the six prominences visible in AIA are also detected in SUVI images. Interestingly, the model detects additional prominences in the SUVI images. Notable in this figure are the effects of the 304~\AA\ channel degradation. Although we have compensated for the image degradation through correction and percentile scaling relative to the first image in the sequence, the residual effects remain. Degradation causes the images to appear dimmer with reduced contrast between prominence and the background. In the following section, we address why the number of predictions varies between these two rows.

\subsection{Dataset Bias}
A key part of our study is highlighting the effects of dataset bias and how they can translate onto model performance. Such biases are not just limited to the spatial distribution of data, but can take the form of temporal biases. This is of importance as machine learning models such as ours are often employed in large-scale statistical studies to study solar cycle variations. We have previously observed signs that degradation may impact performance. Results from Section~\ref{subsec:Subsect3.2} suggest that image degradation may impact model performance, with the number of false positive predictions sharply increasing in more recent, degraded images (see Figure~\ref{fig:Fig2}). The AIA images in the bottom row of Figure~\ref{fig:Fig5} show fewer detections compared to the top row of SUVI images, where there is less degradation present. Therefore, to verify whether there is indeed a relationship with the degradation, we revisit the original dataset distribution.

\begin{figure}[ht]
\graphicspath{ {./images/} }
    \centering
    \includegraphics[width=\linewidth]{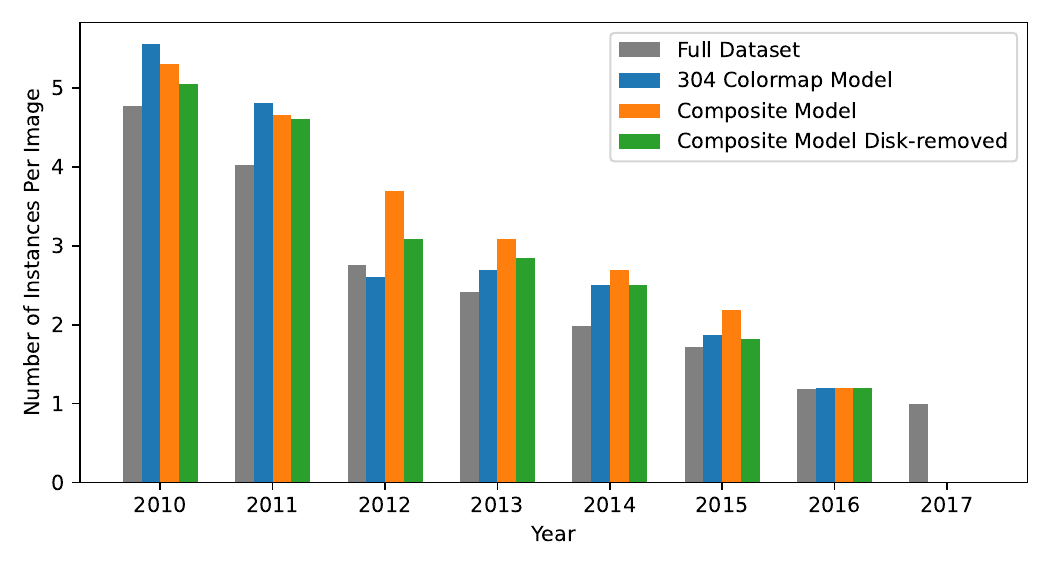} 
    \caption{Number of instances per image for each year, grouped by the full groundtruth labels and model predictions.}
    \label{fig:Fig6}
\end{figure}

Figure~\ref{fig:Fig6} shows the average number of prominence instances per image for each year, comparing ground truth labels with model predictions. The first bar (in gray) shows the ground truth labels from the full dataset. On average, images from 2010 contain the most labels, decreasing with each subsequent year. This is despite 2011 being the most represented year in the dataset, containing $46\%$ of the total labeled prominences. In contrast, 2017 has only 19 labeled prominences. There were no images from 2017 in our testing set. The average temporal cadence for the entire dataset was approximately 22 hours. 

Taken in isolation, the model predictions suggest a correlation between image degradation and detection frequency. However, comparison with the bars for ground truth labels demonstrates that this a direct consequence of the labeling within each image: earlier images have more complete labeling. We speculate that this is due to impaired prominence visibility in later degraded images. Consequently, temporal bias in the labeling has induced an artificial degradation trend in the model predictions. Such effects pervade our processing and could explain the difference in predictions from Figure~\ref{fig:Fig5}.

In addition to the temporal bias, the labeling quality presents additional challenges, as this ultimately underpins the definition of the target class. Inaccurate and inconsistent labeling can impact model performance, and can lead to systematic biases. Accurate labeling of prominences is challenging, as boundaries between nearby structures can be hard to define, especially in the case of discontinuous structures. This is evident when non-maximum suppression fails to suppress multiple bounding boxes on adjacent features, suggesting that the algorithm struggles to determine whether discontinuous structures are related. A potential solution to the issues identified here is to re-label the dataset using processed images, such as the WOW output, which enhances prominence visibility. We leave this task for future work.

\section{Summary} \label{sec:Sect6} 
To summarize, we developed three YOLOv5-based models for automatic prominence detections using the labeled dataset of full-disk SDO/AIA 304~\AA\ prominence observations created by \citet{2021SoPh..296..160B}: the 304 colormap model, the composite model, and the composite disk-removed model. The composite model provides the best performance in detecting prominences, and the best versatility when applied to a different instrument. 

Although the initial 304 colormap model showed strong metric performance, it exhibits a channel-dependent bias, relying predominantly on features encoded in the red channel rather than shape- or texture-based cues across all three channels. This finding motivated the creation of two further composite models which are constructed from three channels: grayscale, WOW, and disk-removed. Such processing enhances prominence structure and contrast, without the arbitrary mapping inherent in colormap selection, and each channel now represents different physical information. These data were additionally calibrated with degradation corrections applied. 

The composite model achieved a recall of $78\%$ and a mAP@50 of 0.749. Visual analysis revealed that both composite models performed better than their metrics suggested, with approximately $74\%$ of apparent false positives were actually valid unlabeled prominences ($80\%$ for the disk-removed variant), resulting in an effective precision of $91\%$ for the composite model and $93\%$ for the disk-removed model. We select the composite model for further analysis due to its exploratory nature in addition to its superior recall, making it better suited for discovering previous uncatalogued prominences. Cross-instrument testing show that the composite models generalizes well to SUVI data. Dataset analysis uncovered patterns and biases that underpin the model predictions, correlating with observation time due to a selection and labeling bias towards earlier images in the SDO/AIA life-cycle. While percentile scaling partially mitigates this issue, complete correction requires relabeling with enhanced processing techniques

Due to the black box nature of deep learning networks, understanding what features are learned and how predictions made is often ambiguous. Therefore, datasets should be curated mindfully of any systematic biases that may transfer into the predictions, both through the image selection and labeling. Based on our findings, we propose guidelines for constructing a more robust dataset for prominences and broader solar physics applications:

\begin{enumerate}
    \item Images should be selected uniformly across the solar cycle to avoid temporal imbalance in the training data.
    \item Images should not contain extraneous features. Unwanted features, such as text, can impair model performance. Preprocessing pipelines should be applied consistently across the entire dataset.
    \item Arbitrary colormaps should be avoided in favor of multi-channel composite data or grayscale representations.
    \item Labeling standards should be consistent. For discontinuous structures, establish clear rules whether to use single or multiple bounding boxes.
    \item Bounding boxes should tightly enclose target features without including irrelevant regions that could confuse the network (e.g., active regions). Features should be centered within the boxes.
\end{enumerate}

Future work will extend the composite model to include further classifications for prominences. As active regions are significantly enhanced in our processing pipeline, we expect that including an active region class label will lead to overall stronger performance for prominence detection as demonstrated in \citet{2024ApJS..272....5Z}. Such a model can be used for statistical studies to investigate prominence variations across the solar cycle. We hope that by combining the model with further processing techniques, such as the prominence RHT method of \citet{2025ApJ...985..161B}, we can obtain further physical properties from the prominences detected by the model. This information can be used to create a prominence catalog that will benefit the solar physics community.

\section*{\textbf{Acknowledgments}}
All authors acknowledge the UK Research and Innovation (UKRI) Science and Technology Facilities Council (STFC) for support from grant No. ST/W006790/1. This study has been supported by the STFC Centre for Doctoral Training in Data Intensive Science (NUdata), as a collaboration between Northumbria and Newcastle Universities. R.J.M. is supported by the UKRI Future Leaders Fellowship (RiPSAW MR/T019891/1 and MR/Z000289/1). The authors would like to thank Sanjiv Tiwari and Meng Jin for their input on the AIA calibration pipeline and degradation correction. SDO/AIA data are courtesy of NASA/SDO and the AIA science team, and were accessed via the Joint Science Operations Center (JSOC) at Stanford University. GOES-18 SUVI data are made available by NOAA's National Centers for Environmental Information (NCEI) through the GOES-R Series program. The authors are willing to provide the data and code upon reasonable request.
\newline
\textit{Software}: PyTorch \citep{paszke2019pytorch}, YOLOv5 \citep{yolov5},
SunPy \citep{2020ApJ...890...68S}, Roboflow \citep{Dwyer2024}. 

\bibliography{references, references2}
\bibliographystyle{aasjournal}

\end{document}